# Simulation of plasma–water interaction with discharge in the existing bubble in water

Niloofar Mohammadi Nahrani[1], *Author*, Maryam Bahreini[2,] *Fellow*, Saeed Hasanpour Tadi[3], *Fellow*

*Abstract*-- **The plasma-liquid interaction is an important issue in plasma technology. The simulation of discharge in spherical bubbles in the water that produced plasma-activated water (PAW) is investigated using finite element methods (FEM) for a simulated 2D dielectric barrier discharge in three different geometries. The electron density changes with voltage, frequency, dielectric thickness, and bubble radius are investigated in different time duration. The results show that electron density increases linearly by increasing voltage, frequency and bubble radius, while it is vice versa for dielectric thickness. High plasma density indicates sufficient plasma-water interaction.**

*Index Terms*-- **bubbles, dielectric barrier discharge, DBD, discharge, plasma activated water, plasma-liquid interaction, simulation.**

## I. Introduction

Plasma refers to a quasi-neutral ionized gas, composed of photons, ions, and free electrons beside excited and neutral atoms[1]. Plasma is divided into thermal and non-thermal (cold plasma). Thermal plasmas belong to thermal equilibrium conditions, which means both electrons and ions are at the same temperature obtained by high energy consumption. When the electron's temperature is much higher than ions, the plasma is non-thermal or cold which does not need as much energy as the thermal one to be obtained[2].

Cold Plasma has a wide range of applications: for example, in medical fields[3],[4],in disinfection and sterilization[5],[6],[7], cancer treatment[8],[9],[10],[11],[12],[13], wound healing[14],[15],[16],[17], and skin rejuvenation[18],[19],[20]. In the Fabric industry plasma is used for strength, water repellency, and fabric dyeability[21]. In the agriculture and food industry, plasma involves in sterilization, pesticide, seed germination, and plant growth[22],[23],[24],[25]. For air purification a nano-oxygen plasma air purification device[26] is used in the investigated inactivation efficiency of bioaerosols. For nanoparticle synthesis, the plasma-generated AgNPs are well-dispersed in the presence of the geminin cationic surfactant and the nano-surfactant compounding system has been improved for antibacterial activity as well as good stability[27]. Another application is material analysis by a plasma discharge, called spark[28],[29].

Furthermore, one of the most important plasma applications is water treatment, which is used in various fields including water purification, bio-sterilization, and decontamination; industrial waste[30],[31], agriculture, and water purification[32],[33],[34].

The interaction of plasma-liquid is becoming an essential topic in the field of plasma technology despite its challenges[29]. Depending on the discharge mechanism, there are different types of plasma-liquid interactions: 1- direct discharges in liquids which are categorized into strong fields discharge like laser pulses (Fig. 1)[35]; 2- discharges in the gas phase over a liquid, including indirect and direct contact with the liquid (Fig. 2)[35], 3- discharges in multiphase environments such as discharges in bubbles inside liquids or discharges contacting liquid sprays or foams. This interaction leads to some instabilities, which produce bubbles and impress the initial micron-sized bubbles (Fig. 3).

The valuable product of this interaction is called plasma-activated water (PAW)[36], which can be produced by one of the above-mentioned electrical discharge mechanisms. In plasma-activated water, reactive oxygen species (ROS) and reactive nitrogen species (RNS) are produced [37]. The RONS in PAW includes long-lived species such as nitrate (NO3-), nitrites (NO2-), hydrogen peroxide (H2O2), and ozone (O3) besides short-lived species such as hydroxyl radicals (OH), nitric oxide (NO), superoxide (O2−), peroxynitrite (OONO2 −) and peroxynitrite (ONOO−) [37]. Due to the production of active species of RONS, PAW can be applied in many applications[38]. RONS are used for cancer cell treatment[39] and facilitation in agricultural applications such as increased seed germination[40]and plant growth by role-playing its species like NO2-, NO3-, H2O2, and NH4. It is also has been used in food industries for disinfection of fruits and vegetables, regulation of enzymes and pesticides in presence of RNS species as well as dental treatment, wound healing and cancer treatment[31],[41].

---

[1]. Niloofar Mohammadi nahrani, master student of Plasma Physics at physics department of Iran *science and technology university*. (e-mail.: nmohammadi111175@gmail.com)

2. Maryam Bahreini, Assistant Professor of physics department at Iran science and technology university. (e-mail: m_Bahreini@iust.ac.ir)

3. Saeed Hasanpour tadi, PhD student at Laser and Plasma Research Institute of Shahid Beheshti University. (e-mail: s_hassanpour@sbu.ac.ir)

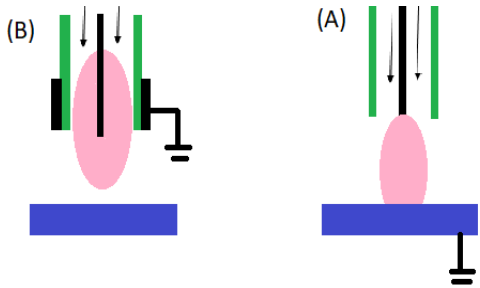

Fig. 1. Indirect liquid discharges; (A) pin to water (B) atmospheric pressure plasma jet (APPJ) geometries

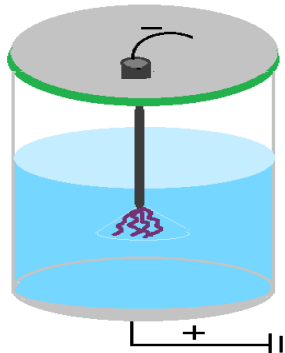

Fig. 2. Direct liquid discharge; an example of the pin-to-plate geometry

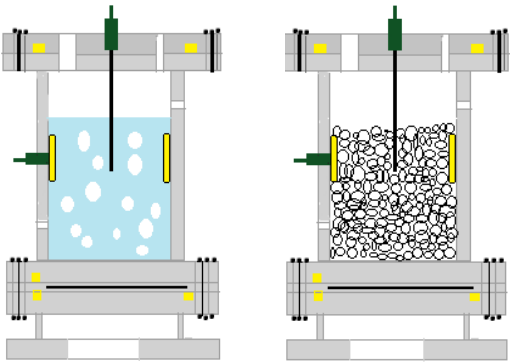

Fig. 3. Multiphase plasma liquid discharge; discharges contacting liquid sprays or foams

The increasing cost of trial and error in experimental research and development procedures leads researchers to use simulation as a powerful method of evaluation. Moreover, investigating and analyzing chemical reactions in electrical discharge through water is sophisticated and expensive. There are some studies in plasma–liquid interaction simulation. N. Vichiansan simulate a plasma jet for NO, OH, and H2O2 production in a simple 2D model and they measure the densities of H2O2 and OH at a distance of 5 mm and 10 mm, respectively[42]. A. Wright simulate a DBD plasma reactor at the gas-liquid interface in 2D asymmetric geometry and investigated the effect of the membrane type on the plasma characteristics and the effect of electrode type on bubble size. In this investigation, two types of steel and nickel layers are considered with the steel layer 1.75 times larger than the nickel layer. The measurement of long-lived species concentrations performed FTIR and UV absorption spectroscopy methods[43]. A. Hamdan and M. S. Cha presented an experimental study of nanosecond discharge in distilled water covered by a layer of dielectric material and simulated the electric field distribution using COMSOL software [40]. Y. Wang used the COMSOL software to model a gas phase surface discharge plasma reactor simulation for yeast inactivation in water and determine the effect of reaction rate coefficient kreac on the initial concentration of yeast cells. They showed that the amount of kreac decreases with increasing the initial concentration[44],[38]. Zhihang Zhao, Xinlao Wei simulated the surface dielectric barrier discharge (SDBD)at low temperature and sub atmospheric pressure. Under a dc voltage of 12 kV, with the propagation of the streamer, the parameters, such as electron density, $N_2^+$ density, $N_2(C)$ density, $O_2^-$ density, electron temperature, space charge density, electric field, and photoionization rate are studied[45].

Direct discharge inside liquids seems to be unfeasible as much power is needed to be applied to achieve suitable results in high voltages and high-power regimes. So, using bubbles inside water let us lead to more achievable results by ordinary discharge devices. In this simulation, electrical discharge inside bubbles has been studied as a noble model. This study simulated electrical discharge in water with the finite element method using COMSOL software by employing the structure of dielectric barrier discharge called DBD during a multiphase discharge considering argon as inlet gas and evaluating the formation of plasma into bubbles. The simulation considers 2D numerical models with a DBD structure and using quartz as a dielectric and the inside of the bubbles filled with argon gas immersed in water. The pressure and temperature are 1 atm and 293 K respectively and the results include electron density distribution, discharge voltage, operating frequency, bubble radius, dielectric thickness, and argon ions.

## II. MATERIAL AND METHOD

*A. Description of modeling geometry*

This simulation considers 2D numerical models with a DBD structure and using quartz as a dielectric with permittivity of 4.5, and the inside of the bubbles filled with argon gas immersed in water. In the geometry, the boundary between the electrode and water is considered a dielectric of glass due to the conservation of charge, and plasma is formed only inside the bubble. The geometry is shown in Fig. 4 that shows a simulated single bubble model with a radius of 0.1 mm.

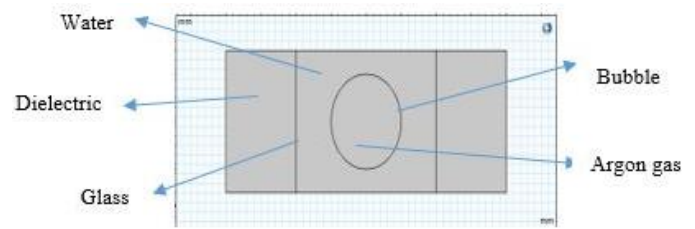

Fig. 4. The schematic diagram of the 2-D simulation model with a single bubble.

## B. Governing Equations

The numerical model of the barrier discharge solves a pair of drift-diffusion equations for the electron density and mean electron energy of (1) and (2), and the heavy species transport equation for non-electron species (3). The electrostatic field is computed using Poisson equation (3) by the finite element method using COMSOL Multi-physics software:

$$\frac{\partial}{\partial t}(n_e) + \nabla \cdot [-n_e(\mu_e \cdot E) - D_e \nabla n_e] = R_e \quad (1)$$

$$\frac{\partial}{\partial t}(n_\varepsilon) + \nabla \cdot [-n_\varepsilon(\mu_\varepsilon \cdot E) - D_\varepsilon \nabla n_\varepsilon] + E \cdot \Gamma_e = R_\varepsilon \quad (2)$$

The electron source is Re and the energy loss due to inelastic collisions is Rε. The electron diffusivity, energy mobility, and energy diffusivity are computed from the electron mobility using:

$$D_e = \mu_e T_e \;,\; \mu_\varepsilon = (\frac{5}{3})\mu_e \;,\; D_\varepsilon = \mu_\varepsilon T_e \quad (3)$$

The source coefficient in the above equations is determined by plasma chemistry using rate coefficients.

For non-electron species (4):

$$\rho \frac{\partial}{\partial t}(w_k) + \rho(u.\nabla)w_k = \nabla \cdot j_k + R_k \quad (4)$$

the electrostatic field is (5):

$$-\nabla \cdot \varepsilon_0 \varepsilon_r \nabla V = \rho \quad (5)$$

The space charge density ρ is automatically computed based on the plasma chemistry specified in the model using the following formula (6):

$$p = q\left(\sum_{k=1}^{N} Z_k n_k - n_e\right) \quad (6)$$

## C. Boundary Conditions

Electrons are lost to the wall due to random motion within a few mean free paths of the wall and gained due to secondary emission effects, resulting in the following boundary condition for the electron flux (7):

$$n.\Gamma_e = \left(\frac{1}{2}V_{e,th}n_e\right) - \sum_p \gamma_p(\Gamma_p.n) \quad (7)$$

The electron energy flux is (8):

$$n.\Gamma_\varepsilon = \left(\frac{5}{6}V_{e,th}n_\varepsilon\right) - \sum_p \varepsilon_p \gamma_p(\Gamma_p.n) \quad (8)$$

The second term on the right-hand side of Equation (7) is the gain of electrons due to secondary emission effects with γp being the secondary emission coefficient. The second term in Equation (8) is the secondary emission energy flux with εp being the mean energy of the secondary electrons.

The set of reactions and collisions considered in the simulated model are given in Table I.

TABLE I
Table of collisions and chemical reactions in the simulated model

| REACTION | FORMULA | TYPE | Δε(eV) |
|---|---|---|---|
| 1 | e+Ar=>e+Ar | Elastic | 0 |
| 2 | e+Ar=>e+Ars | Excitation | 11.5 |
| 3 | e+Ars=>e+Ar | Superplastic | -11.5 |
| 4 | e+Ar=>2e+Ar+ | Ionization | 15.8 |
| 5 | e+Ars=>2e+Ar+ | Ionization | 4.28 |
| 6 | Ars+Ars=>e+Ar+Ar+ | Penning ionization | - |
| 7 | Ars+Ar=> Ar+Ar | Metastable quenching | - |

## III. RESULTS AND DISCUSSION

In this simulation, one of the electrodes is assumed to be ground and a sinusoidal voltage is applied to the other electrode. The initial electron density is assumed to be $10^{13}$ m$^{-3}$. We investigate the effect of changes in parameters of frequency, voltage, dielectric thickness, and bubble radius on electron density.

### A. Voltage

The numerical Figs. 5, 6, and 7 show the dynamic behavior of electron density over time at different voltages. The simulation was performed with constant conditions of 50 kHz frequency, bubble radius of 0.1 mm, and gap distance of 1 mm. Fig. 8 illustrates the results of electron density for voltage of 1000 volts, which at the first time (t=0 microsecond) initial electron density is (a) $10^{13}$ m$^{-3}$ and at last time (t=20 microsecond) electron density is equal to (j) $10^{10}$ m$^{-3}$. The behavior of the electron density over time by varying voltages from 500 to 1000 volts was the as seen in Fig. 5, non-uniform electron density changes observed as same as Wenjun Ning, Mark J Kushner and et al. by 20 Kv applied voltage and 4mm of bubble diameter for times= 0.7,1,1.5,1.9,2.5,4.4,5.6.10 ns[46]. Figs. 6 and 7 show the results of electron density for voltages of 1500 and 2000 volts respectively, which at t=20 microsecond electron density is (l) $10^{16}$ m$^{-3}$ and (j) $10^{20}$ m$^{-3}$. A direct relation between time and density as Dmitry Levko, Ashish Sharma and Laxminarayan L Raja has done before in the shorter period of time (t=1,2,3.6,4.4 ns) by applying voltage of -9Kv Dc[47], and Juliusz et al. by applying -13 Kv Dc, 0.1mm of bubble diameter, for times= 0.4,0.5,0.6,1.7 ns[48]. Applying a voltage to the electrodes creates an electric field between them and as the voltage increases, the electric field increases. So, the collision energy between particles would be more than before. Therefore, more ionization leads to a developing plasma region and more free electron production. The result shows that the electron density increases linearly by changing voltage. In this study microsecond period employed because of low applied voltage in comparison with the same named works, as nano-second periods weren't enough time to initiating plasma; No electron density increasing detected in short nano-second period.

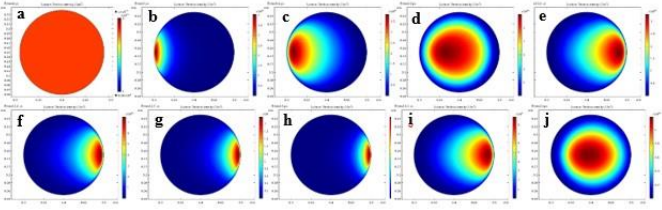

Fig. 5. Electron density changes by passing time in different region of bubble for 1000 applied voltage; Electron density at t=0us was $10^{13}$ m$^{-3}$ (a) at t=5 us was $4 \times 10^{12}$ m$^{-3}$ (b), at t=9.4 us was $3.6 \times 10^{10}$ m$^{-3}$ (c), at t=10 us was $5.7 \times 10^{10}$ m$^{-3}$ (d), at t=10.6 us was $3.7 \times 10^{10}$ m$^{-3}$ I, at t=11.6 us was $7.5 \times 10^{10}$ m$^{-3}$ (f), at t=12.5 us was $2.5 \times 10^{11}$ m$^{-3}$ (g), at t=15 us was $9 \times 10^{11}$ m$^{-3}$ (h), at t=19.2 us was $8.3 \times 10^{9}$ m$^{-3}$ (i), at t=20 us was $1.8 \times 10^{10}$ m$^{-3}$ (j)

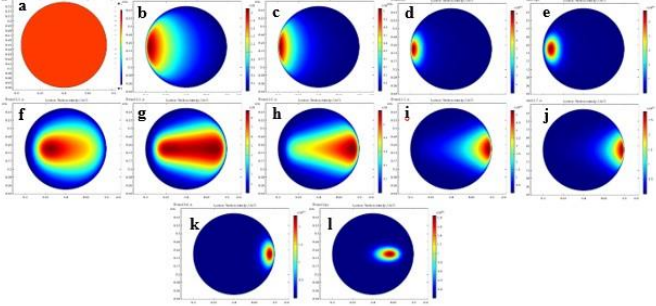

Fig. 6. Electron density changes by passing time in different region of bubble for 1500 applied voltage; Electron density at t=0 us was $10^{13}$ m$^{-3}$ (a) at t=0.7 us was $2 \times 10^{11}$ m$^{-3}$ (b), at t=1.5 us was $1 \times 10^{12}$ m$^{-3}$ (c), at t=9 us was $9 \times 10^{11}$ m$^{-3}$ (d), at t=10 us was $5 \times 10^{12}$ m$^{-3}$ I, at t=10.1 us was $1.2 \times 10^{12}$ m$^{-3}$ (f), at t=10.3 us was $8 \times 10^{11}$ m$^{-3}$ (g), at t=10.5 us was $1 \times 10^{12}$ m$^{-3}$ (h), at t=11.1 us was $5 \times 10^{12}$ m$^{-3}$ (i) at t=11.7 us was $2.5 \times 10^{13}$ m$^{-3}$ (j) at t=19.8 us was $2 \times 10^{14}$ m$^{-3}$ (k) at t=20 us was $1.8 \times 10^{16}$ m$^{-3}$ (l)

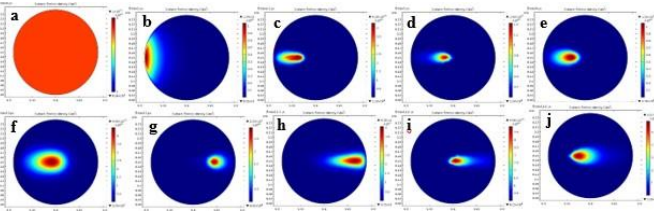

Fig. 7. Electron density changes by passing time in different region of bubble for 2000 applied voltage; Electron density at t=0 us was $10^{13}$ m$^{-3}$ (a) at t=2 us was $1.04 \times 10^{13}$ m$^{-3}$ (b), at t=4.1 us was $4.29 \times 10^{19}$ m$^{-3}$ (c), at t=5 us was $1.42 \times 10^{20}$ m$^{-3}$ (d), at t=9 us was $1.05 \times 10^{19}$ m$^{-3}$ I, at t=10 us was $4.48 \times 10^{19}$ m$^{-3}$ (f), at t=11 us was $2.93 \times 10^{18}$ m$^{-3}$ (g), at t=11.1 us was $4.06 \times 10^{17}$ m$^{-3}$ (h), at t=12.6 us was $4.86 \times 10^{19}$ m$^{-3}$ (i), at t=14.6 us was $5.87 \times 10^{20}$ m$^{-3}$ (j)

### B. Frequency

The frequency range was from 20 kHz to 80 kHz with constant conditions of the voltage of 1500 volts, bubble radius of 0.1 mm, and gap distance of 1 mm. Fig. 8 shows the results of electron density for that in the frequency of 80 kHz last time (t=16 microsecond) the electron density value is (O)$10^{20}$m$^{-3}$. The result shows that as frequency increases, the electron density increases. By changing electric field direction, electrons would change their direction too. The power source as the controller of this procedure by a parameter called frequency could limit the electron's path. This limitation causes more electron trapping and more electron density inside the bubble.

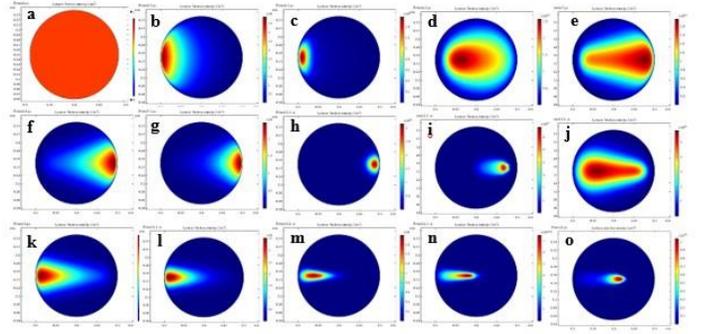

Fig. 8. Electron density changes by passing time in different region of bubble for 1500 volts applied voltage and frequency=80KHz; Electron density at t=0 us was $10^{13}$ m$^{-3}$ (a) at t=0.7 us was $4.5 \times 10^{11}$ m$^{-3}$ (b), at t=6.1 us was $1.4 \times 10^{12}$ m$^{-3}$ (c), at t=6.3 us was $2.5 \times 10^{12}$ m$^{-3}$ (d), at t=6.5 us was $1.8 \times 10^{12}$ m$^{-3}$ I, at t=6.8us was $6 \times 10^{12}$ m$^{-3}$ (f), at t=7.1 us was $2 \times 10^{13}$ m$^{-3}$ (g), at t=12.3 us was $3 \times 10^{15}$ m$^{-3}$ (h), at t=12.5 us was $6 \times 10^{17}$ m$^{-3}$ (i), at t=12.6 us was $5 \times 10^{15}$ m$^{-3}$ (j), at t=13 us was $8 \times 10^{15}$ m$^{-3}$ (k), at t=13.2 us was $3.5 \times 10^{16}$ m$^{-3}$ (l), at t=13.8 us was $1.6 \times 10^{18}$ m$^{-3}$ (m), at t=14.2 us was $7 \times 10^{18}$ m$^{-3}$ (n), at t=16 us was $1 \times 10^{20}$ m$^{-3}$ (o),

### C. Dielectric thickness

The In this part of the simulation, the dielectric thickness is considered to be 0.5 mm and 1.5 mm with constant conditions of the voltage of 1800 volts, frequency of 50 kHz, and bubble radius of 0.1 mm. Figs. 9 and 10 show the results of electron density of $10^{20}$ m$^{-3}$ in dielectric thickness of 0.5 and the electron density of $10^{15}$ m$^{-3}$ in dielectric thickness of 1.5 mm. The result illustrates the electron density decreases with increasing dielectric thickness. Increasing dielectric thickness would decrease the capacitance of the system so lower electrical charges on electrodes and lower electric field lead to less electron density.

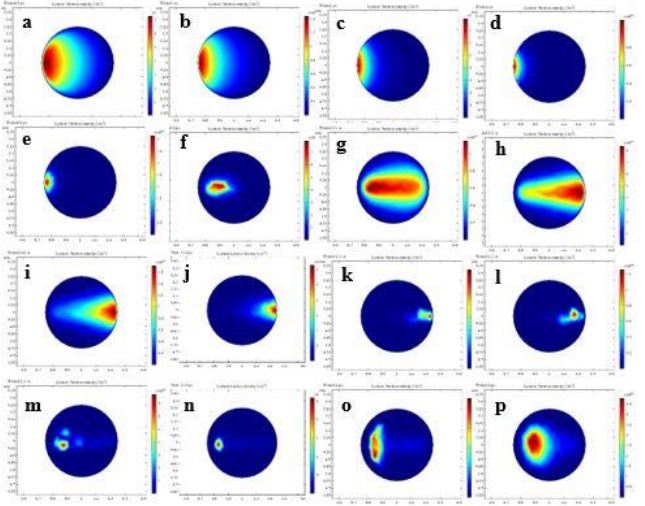

Fig. 9. Electron density changes by passing time in different region of bubble for dielectric thickness=0.5mm; Electron density at t=0.5 us was $7 \times 10^{11}$ m$^{-3}$ (a) at t=1 us was $1.4 \times 10^{12}$ m$^{-3}$ (b), at t=2 us was $1.4 \times 10^{13}$ m$^{-3}$ (c), at t=4 us was $1 \times 10^{15}$ m$^{-3}$ (d), at t=9.9 us was $4 \times 10^{13}$ m$^{-3}$ I, at t=10 us was $7 \times 10^{14}$ m$^{-3}$ (f), at t=10.1 us was $1 \times 10^{14}$ m$^{-3}$ (g), at t=10.5 us was $8 \times 10^{13}$ m$^{-3}$ (h), at t=10.8 us was $1.6 \times 10^{14}$ m$^{-3}$ (i), at t=11.6 us was $3 \times 10^{15}$ m$^{-3}$ (j), at t=12.1 us was $1.4 \times 10^{17}$ m$^{-3}$ (k), at t=12.2 us was $1.2 \times 10^{18}$ m$^{-3}$ (l), at t=12.3 us was $7 \times 10^{19}$ m$^{-3}$ (m), at t=12.5 us was $2.5 \times 10^{20}$ m$^{-3}$ (n), at t=14 us was $4.5 \times 10^{20}$ m$^{-3}$ (o), at t=20 us was $4.5 \times 10^{20}$ m$^{-3}$ (p)5

Fig 10. Electron density changes by passing time in different region of bubble for dielectric thickness=1.5mm; Electron density at t=0.4 us was $7\times10^{11}$ m$^{-3}$ (a) at t=1.9 us was $5\times10^{12}$ m$^{-3}$ (b), at t=4.3 us was $1.8\times10^{14}$ m$^{-3}$ (c), at t=9.9 us was $1\times10$

### D. Bubble radius

In this part, the bubble size is considered to be 0.06- and 0.09-mm. Figs. 11 and 12 show that increasing the radius of the bubble leads to an increase in electron density.

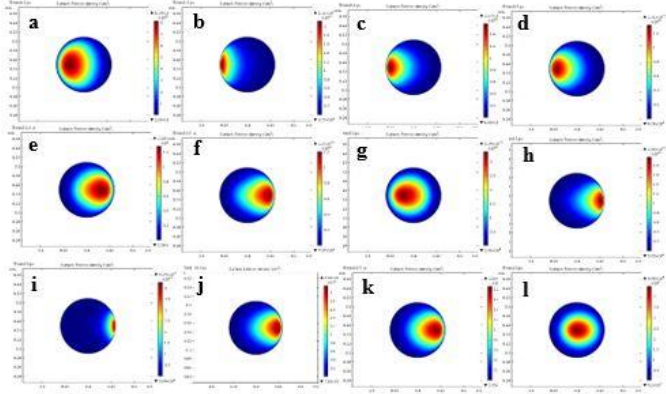

Fig11. Electron density changes by passing time in different region of bubble for radius =0.06mm; Electron density at t=0.8 us was $8.21\times10^{10}$ m$^{-3}$ (a), at t=1.5 us was $2.11\times10^{11}$ m$^{-3}$ (b), at t=8 us was $1.24\times10^{11}$ m$^{-3}$ (c), at t=9.3 us was $1.57\times10^{10}$ m$^{-3}$ (d), at t=9.7us was $1.31\times10^{10}$ m$^{-3}$ I, at t=10 us was $3.34\times10^{10}$ m$^{-3}$ (f), at t=10.5 us was $1.2\times10^{10}$ m$^{-3}$ (g), at t=11.5 us was $4.22\times10^{10}$ m$^{-3}$ (h), at t=18.6 us was $7.54\times10^{9}$ m$^{-3}$ (i), at t=19.4 us was $2.16\times10^{9}$ m$^{-3}$ (j), at t=19.7 us was $1.85\times10^{9}$ m$^{-3}$ (k), at t=20 us was $4.46\times10^{9}$ m$^{-3}$ (l).

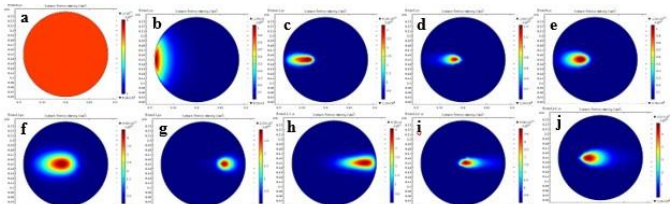

Fig 12. Electron density changes by passing time in different region of bubble for radius = 0.09 mm; Electron density at t=0 us was $10^{13}$ m$^{-3}$ (a) at t=2 us was $1.04\times10^{13}$ m$^{-3}$ (b), at t=4.1 us was $4.29\times10^{19}$ m$^{-3}$ (c), at t=5 us was $1.42\times10^{20}$ m$^{-3}$ (d), at t=9 us was $1.05\times10^{19}$ m$^{-3}$ I, at t=10 us was $4.48\times10^{19}$ m$^{-3}$ (f), at t=11 us was $2.93\times10^{18}$ m$^{-3}$ (g), at t=11.1 us was $4.06\times10^{17}$ m$^{-3}$ (h), at t=12.6 us was $4.86\times10^{19}$ m$^{-3}$ (i), at t=14.6 us was $5.87\times10^{20}$ m$^{-3}$ (j)

The electron density distribution of excited atoms is shown in Fig. 13. The excited atoms are different from the electron den-sity. They can return to the ground state and emit light. Hence this pattern is plasma eyesight which can be observed with a camera or eye.

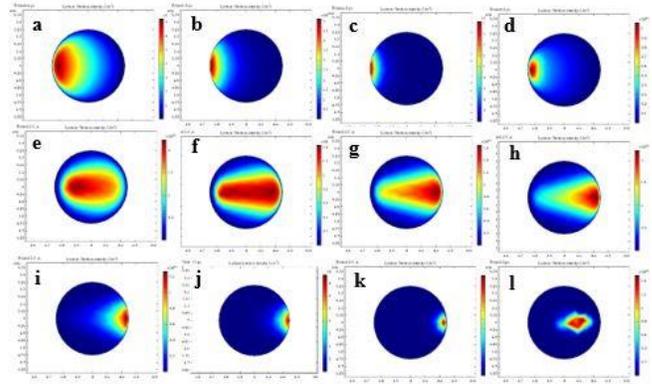

Fig 13. Electron density distribution of excited atom.

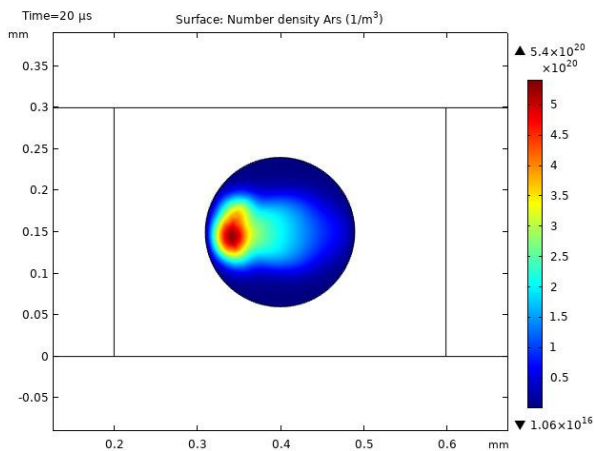

## III. CONCLUSION

The two-dimensional simulation of a single bubble with a DBD plasma structure is studied using COMSOL software. The effect of changes in some parameters including voltage, frequency, dielectric thickness, and bubble radius on electron density has been studied. The results show that increasing voltage, frequency, and bubble radius have a direct effect on increasing the electron density. Electron density as a sign of plasma region evidently has been changed by these parameters. So, forming a plasma in a bubble inside water is possible by selecting the appropriate frequency and voltage and it is possible for micro-size bubbles more than nano-size bubbles. Creating a high voltage difference between a nanometer-diameter sphere is more difficult than a micrometer bubble and requires a stronger generator due to Paschen's curve which says as the bubble size becomes smaller, the formation of the discharge requires a higher voltage.

## IV. REFRENCES